# Influence of temperature and humidity on the detection of benzene vapor by piezoelectric crystal sensor

Chan-Hyon Han[1], Yong-A Choe[1], Jong-Ho Yun[1], Kye-Ryong Sin[2,*]

[1] *Analytical Research Insutitute, **Kim Il Sung** University*

[2]*Faculty of Chemistry, **Kim Il Sung** University*

*Taesong District, Pyongyang, Democratic People's Republic of Korea*

Email Address: ryongnam9@yahoo.com

**Abstract:**

The effects of temperature and humidity on the estimation of air pollution by benzene by using the piezoelectric crystal gas sensor were studied. Polyvinylchloride films were used as substrate for the immobilization of polymethylphenylsiloxane onto the electrode surface of the piezoelectric crystal. The sensing layer consisting of polymethylphenylsiloxane and polyvinylchloride was used for real-time monitoring of benzene, one of the atmospheric pollutants. According to the humidity from 35% to 75%, the upper limit of detection by this sensor was decreased and the response time and frequency recovery time for detecting benzene were long. On the other hand, as increasing the temperature from 5℃ to 60℃, the response time and the frequency recovery time of the sensor were short, but its sensitivity got worse. The models for the correlation between the benzene concentration and temperature (or humidity) were presented.

## 1. Introduction

The analysis of volatile organic compounds are currently performed by using standard analytical instruments such as gas chromatography and mass spectrometry, but these analytical instruments are not suitable for real-time monitoring and cannot be used as small portable devices. Chemical sensor technology can provide low-cost and small size devices, which are desirable for a wide range of application such as environmental monitoring and industrial process analysis. Piezoelectric quartz crystal (PQC) or the surface acoustic wave (SAW) oscillators with a chemical sensing layer have been developed into highly sensitive devices for the detection of traces of atmospheric pollutants. The sensing layer on the surface of PQC is a critical component for a mass-sensitive sensor and responsible for generating the analytical signal from the interaction between the analytes and the sensing layer.

The piezoelectric chemical gas sensor has been used in *in-situ* detection [1-4, 7, 9, 16, 18, 21] and for real-time monitoring [5, 6, 8, 10-15, 17, 19, 20, 22] for the atmospheric pollutants. The temperature and humidity are important factors for sensing the pollutants under the atmospheric conditions. Therefore, the effects of temperature and humidity on the sensing must be considered for the rapid and correct detections of the pollutants.

In this paper, temperature and humidity influence on the detection of benzene vapors in the atmosphere by the piezoelectric crystal sensor was studied.

## 2. Experimental

### 2.1. Apparatus and reagents

The fabricated silver (diameter 6mm) was vacuum-deposited on both sides of mechanically polished AT-cut 4MHz quartz crystal (diameter 12mm, International Crystal Manufacturing Co. Inc., Oklahoma City, OK). The lead wires of quartz crystal were connected to a self-constructed IC-TTL oscillator. Both of the detector cell and the oscillator were put into a copper Faradic cage to remove interference from the surrounding electromagnetic noises. The frequency was monitored by a universal frequency counter (HP53131A), and the data recorded were stored in a computer.

All chemicals such as polymethylphenylsiloxane (OV-1701), polyvinylchloride, tetrahydrofuran, and benzene were analytical reagent grade.

**2.2. Procedure**

**2.2.1. Manufacture of the piezoelectric chemical sensor**

After the silver-plated quartz crystals were treated with 0.1M NaOH for 5 min and washed with distilled water, acetone, and distilled water again, and dried in air.

The coating solution was prepared by putting 10mg of polymethylphenylsiloxane and 50mg of polyvinylchloride into 5mL tetrahydrofuran.

The quartz plate electrode with resonance frequency of 4MHz was covered with the prepared coating solution by using the micro-injector. This piezoelectric crystal gas sensor, hereafter, was called the crystal senor in short.

**2.2.2. Examination of characteristics of the piezoelectric crystal gas sensor**

The organic vapor samples were generated by bubbling nitrogen gas through a target organic solvent to produce its vapor for testing. The organic vapor was subsequently diluted with the carrier gas in several steps to produce desired concentration. The dilution factor was controlled through adjusting the flow rates of both of the organic vapors and the carrier gas by flow controllers.

The crystal sensor was placed in a stainless steel test chamber (2cm×3cm×3cm). At first, the carrier stream of dried nitrogen (4mL/s) was passed through the test cell. When the frequency signal became stable, the sample stream containing benzene vapor was introduced into the detection cell. The test gas was prepared by mixing the diluted organic vapor stream (0.8mL/s) and the dried nitrogen stream (4mL/s) with the desired proportions.

Benzene sensing characteristics of the crystal sensor were measured in the different temperatures and humidities.

**3. Results and Discussion**

**3.1. Effects of Temperature**

The response time of the crystal sensor is the one of the important factors on the rapidity of the detection by the sensor. At the fixed humidity (60%), the response time of the crystal sensor to benzene was measured in the different temperatures. As shown in Figure 1, with increasing temperature (5¬60℃), the response time of the crystal sensor became decreased, but above 30℃, its change became slow.

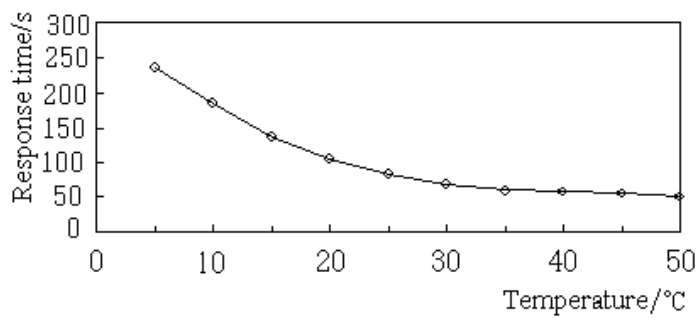

Fig. 1    Changes of the response time according to the temperature

The frequency recovery time of the crystal sensor is the main factor for the real-time detection of the sample. The frequency was measured in the detecting cell with the given concentration of benzene, and then the time for recovery to the initial frequency by the air exchange were measured in the different temperatures. Figure 2 shows that as increasing the temperature, the frequency recovery time became short, which means the rapid desorption of benzene adsorbed on the sensor with increasing temperature.

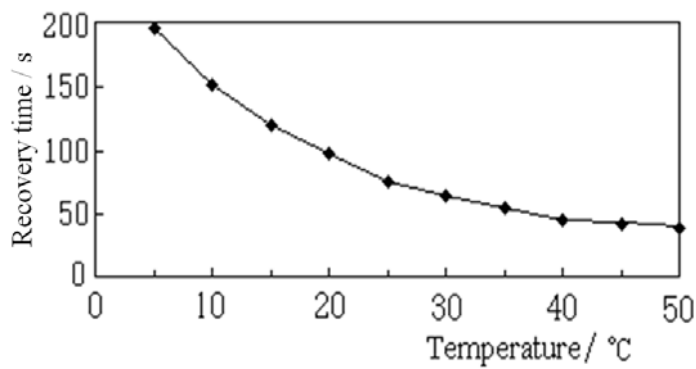

Fig. 2    Changes of the frequency recovery time according to the temperature

The frequency change (△f) according to the temperature under the given concentration of benzene (1.2±0.1μg/mL) was shown in Figure 3. As increasing the temperature, the amount of the frequency change became small.

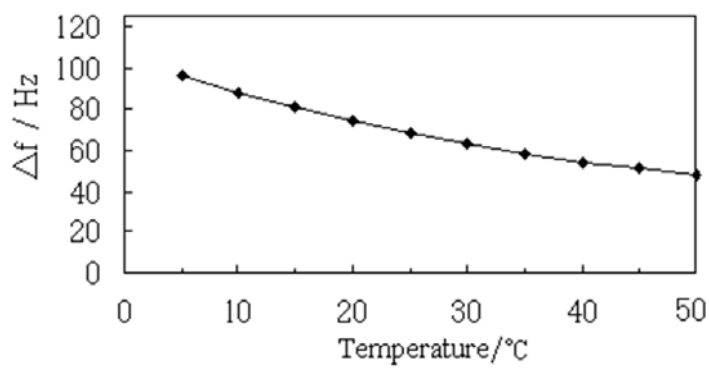

Fig. 3    Effect of the temperature on the frequency change

From the above results, it is clear that the temperature increase is good for reducing the response time or frequency recovery time, but the detecting sensibility of the crystal sensor became worse.

With changing the benzene concentration ($C_B$) and the detecting temperature, the amount of frequency change of the crystal sensor was measured. (Fig. 4)

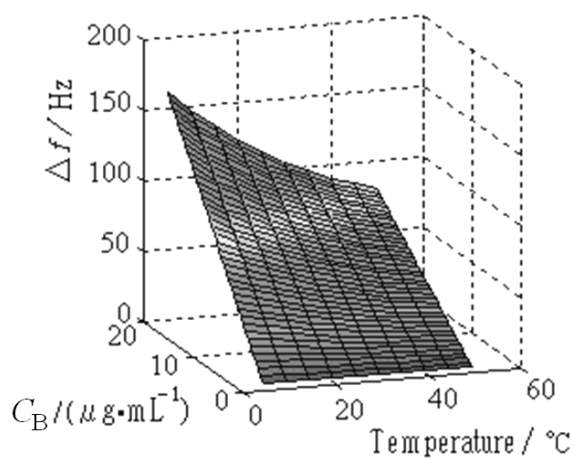

Fig. 4    Frequency change according to the benzene concentration ($C_B$) and temperature.

For determination of the benzene concentration from the amount of frequency change of the sensor by considering the temperature effect, one model equation was presented like Eq. 1.

$$\Delta f = C_B(AT^2 + BT + C) + DT^2 + ET + F \tag{1}$$

where $\Delta f$ is the amounts of frequency change (Hz), $C_B$ is the concentration of benzene (μg/mL), $T$ is the temperature (℃) and $A$, $B$, $C$, $D$, $E$, $F$ are constants.

By using the least square method, the regression coefficients ($A$, $B$, $C$, $D$, $E$, $F$) were determined from the measured data and found out that coefficients D, E, F were below $10^{-7}$, which can be ignored hereafter. Therefore, the benzene concentration can be determined from Eq. 2.

$$C_B = \frac{\Delta f}{AT^2 + BT + C} \tag{2}$$

where A=0.0125Hz・mL/(μg・K$^2$), B=-1.561Hz・mL/(μg・K) and C=87.98Hz・mL/μg.

**3.2. Effects of the humidity**

Under the fixed temperature (20±0.5℃) and different humidity (35¬75%), the frequency change according to the benzene concentration was measured by the crystal sensor. (Fig. 5)

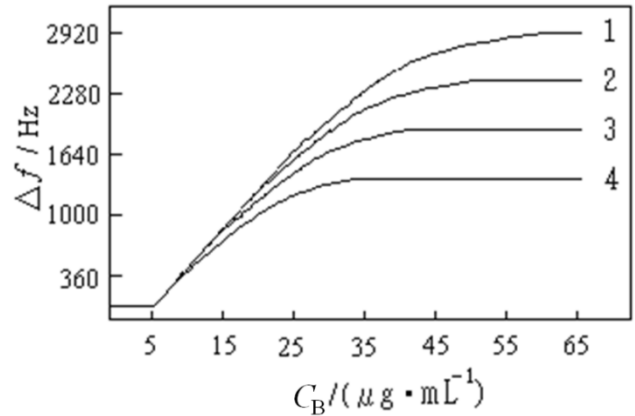

Fig. 5    Frequency change according to the benzene concentration

(humidity: 1- 64%, 2- 66%, 3- 68%, 4- 70%)

As shown in Figure 5, the linear range became narrow and the change in the upper boundary of the detection was increased according to the humidity increase.

The changes of the linear range and the upper boundary of the detection of the sensor according to the humidity were dependent on the polarity of the sample. Figure 6 showed the frequency change ($\Delta$f) according to the concentration of benzene (non-polar) by the crystal sensor and to the concentration of acetone (polar) by the sensor coated with PEG-1500. As shown in Figure 6, the △f width (△w) was small in the crystal sensor with high sensibility on the non-polar sample (benzene), large in the PEG-1500 coated sensor with high sensibility on the polar sample (acetone).

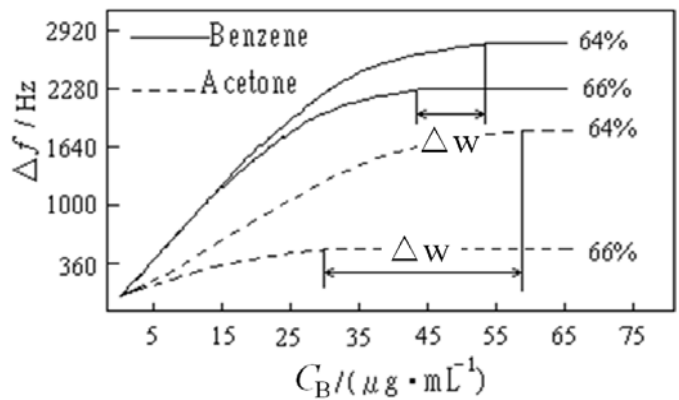

Fig. 6.   Frequency changes according to the concentrations of benzene and acetone

The response time and frequency recovery time were also measured under the different humidity (Fig. 7 and Fig. 8).

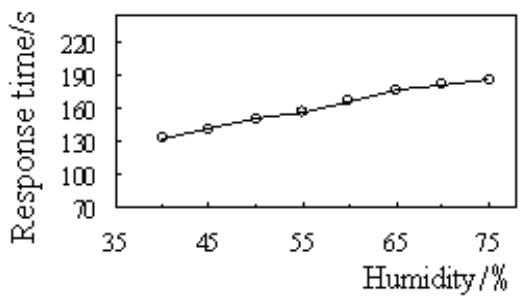

Fig. 7   Response time change according to the humidity

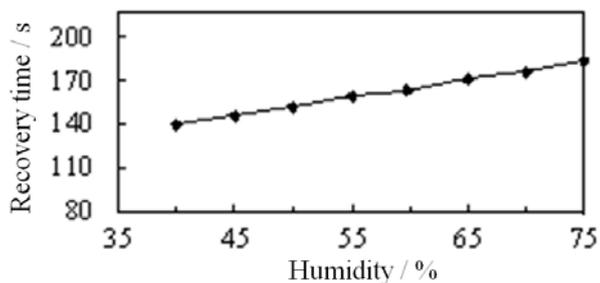

Fig. 8.   Change of the frequency recovery time according to the humidity

From Figure 7 and 8, it can be seen that the high humidity increased the response time and the frequency recovery time of the crystal sensor. This is related to the slow desorption of the moisture adsorbed on the sensing sample.

Therefore, in the detection of the volatile organic vapor, the amount of moisture in the detected substances must be reduced, or the effects of moisture must be considered in the measured detection values.

Under the different humidity and the fixed benzene concentration (1.2±0.1μg/mL), the frequency changes were measured (Fig. 9).

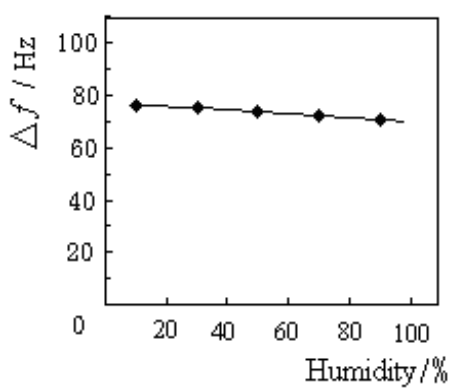

Fig. 9    Frequency change according to the humidity

Figure 9 showed that high humidity decreased the sensibility of the crystal sensor owing to the moisture adsorbed on benzene. To correct the effect of humidity in determination of benzene concentration, presented was a model equation (3) as following:

$$C_B = \frac{\Delta f}{Ah + B} \quad (3)$$

where $h$ was humidity (%), A=-0.2321Hz·mL/(μg·%) and B=75.124 Hz·mL/μg, which were determined by the same routine for Eq. 2.

## 4. Conclusion

The effects of the temperature and humidity on the benzene concentration in the atmosphere measured by using the piezoelectric crystal sensor coated with polymethylphenylsiloxane were discussed and two model equations were presented for determining the benzene concentration under the different temperature and humidity.